\documentclass[prl,numerical,preprint,showkeys,longbibliography]{revtex4-1}

\usepackage{epsfig,amsmath,amssymb,txfonts,hyperref} 
\usepackage{pdfpages} 

\makeatletter 
\AtBeginDocument{\let\LS@rot\@undefined}
\makeatother

\newlength{\wholefigwidth}
\newlength{\smallfigwidth}
\newlength{\halfsmallfigwidth}
\newlength{\figwidth}
\begin{document} 

\title{Quantifying the contributions to diffusion in complex materials}

\author{Soham Chattopadhyay}
\author{Dallas R. Trinkle}
\email{dtrinkle@illinois.edu}
\affiliation{Department of Materials Science and Engineering, University of Illinois, Urbana-Champaign, Illinois 61801, USA}

\date{\today}

\begin{abstract}
  Using machine learning with a variational formula for diffusivity, we recast diffusion as a sum of individual contributions to diffusion---called ``kinosons''---and compute their statistical distribution to model a complex multicomponent alloy. Calculating kinosons is orders of magnitude more efficient than computing whole trajectories, and elucidates kinetic mechanisms for diffusion. The density of kinosons with temperature leads to new accurate analytic models for macroscale diffusivity. This combination of machine learning with diffusion theory promises insight into other complex materials.
\end{abstract}
\keywords{diffusion; mass transport; variational principle; machine learning}

\maketitle

The first verified laws for diffusion in fluids dates back to the nineteenth century work of Fick\cite{Fick1855}, and the last century brought Einstein's major breakthrough connecting Brownian motion and diffusion\cite{Einstein1905}, and Onsager's nonequilibrium thermodynamics\cite{Onsager1931}. In solids, systematic studies of diffusion in metals go back to Roberts-Austen work on gold diffusing into lead\cite{RobertsAusten1896}. As the fundamental kinetic process for atomic motion in a material, diffusion controls materials processing for metals, semiconductors, ceramics, and nanoparticles; the operation of batteries and fuel cells; and degredation from corrosion and irradiation\cite{Balluffi-Kinetics}. The nano- or atomic-scale processes controlling diffusion are often thermally activated and driven by external forces; the understanding of thermally activated processes goes back to Arrhenius\cite{Arrhenius1889}, Eyring\cite{Eyring1935}, Polanyi\cite{Polanyi1935}, and Vineyard\cite{Vineyard1957}. However, in complex materials with multiple competing processes, diffusivity can deviate from Arrhenius behavior, and is difficult to connect individual processes to macroscale behavior.

Theoretical approaches to diffusion abound, where a variety of approximations have been developed\cite{Heitjans2005,Mehrer2007}.
An incomplete list of approaches includes stochastic methods like kinetic Monte Carlo\cite{Murch1984,Belova2000,Belova2001,Belova2003a,Belova2003b,Athnes2022}, master-equation methods based on cluster expansions\cite{Nastar2000,Nastar2005,Schuler2020} and kinetic mean-field approximations\cite{Belashchenko1998,Vaks2014,Vaks2016}, path probability methods for irreversible thermodynamics\cite{Kikuchi1966,Sato1983,Sato1985}, Green function methods\cite{Montroll1965,Koiwa1983,TrinkleOnsager2017}, and Ritz variational methods\cite{Gortel2004,ZaluskaKotur2007,ZaluskaKotur2014}. Recent work on a variational method\cite{TrinkleVariational2018} connected many of these methods, and provided a basis for comparing accuracy. The computational methods take different approaches to the underlying difficulty in diffusion: the long time limit of trajectories complicates identifying important processes and obscures how individual states and transitions contribute to transport. Complex systems have a variety of rates that themselves may be Arrhenius, but together the diffusivity deviates from Arrhenius behavior. A state of the system may have fast transitions to other states, but without connected pathways these fast transitions indicate trapping without contributing to diffusion, leading to emergent behavior such as percolation.

The variational principle for diffusion combined with modern machine learning techniques presents an opportunity: by rewriting the diffusivity of a system as a sum over individual contributions from every state and transition---what we call ``kinosons'' (corresponding to movement)---we analyze the transport processes in a new way, and discover non-Arrhenius behavior in complex systems. Machine learning methods solve the optimization problem in the variational method to accelerates the computation of diffusivity and permit the decomposition into kinosons. With the density of kinosons, we directly identify new analytic forms for the diffusivity and identify differences in behavior of species within a system. We demonstrate these ideas with diffusion in a complex high entropy alloy, find fingerprints of percolation, and a new analytic form for diffusion.

Diffusivity can be alternatively computed from the mean squared displacement at infinite time, or as the minimum of average squared displacements. The Einstein-Smoluchowski form\cite{Einstein1905} of diffusion,
\begin{equation}
  D = \lim_{t\to\infty}\frac{
    \left\langle
    \left(\mathbf{x}(t) - \mathbf{x}(0)\right)^2
    \right\rangle}%
  {2dt}
  \label{eqn:Einstein}
\end{equation}
expresses diffusivity $D$ in $d$-dimensions as the average of the long-time limit of squared displacement $\mathbf{x}(t)-\mathbf{x}(0)$ divided by time. By contrast, for a Markovian system, a variational form\cite{TrinkleVariational2018} for Onsager transport coefficients,
\begin{equation}
  D = \inf_{\mathbf{y}_\chi}\frac1{2d}
  \Bigg\langle
  \sum_{\chi'} W(\chi\to\chi')\left(\mathbf{\delta x}(\chi\to\chi') +
  \mathbf{y}_{\chi'} - \mathbf{y}_{\chi}\right)^2
  \Bigg\rangle_\chi
  \label{eqn:variation}
\end{equation}
is as an average over contributions from every transition between any pair of states $\chi$ and $\chi'$, with rate $W(\chi\to\chi')$ and displacement $\mathbf{\delta x}(\chi\to \chi')$. The diffusivity is minimized by optimizing the ``positions'' of states $\mathbf{y}_\chi$. If every state is moved to its optimal position $\mathbf{y}_s$, then the mean displacement out of every state is zero, while the mean squared displacement grows linearly with time. The optimized displacements between states is $\widetilde{\mathbf{\delta x}}(\chi\to\chi') := \mathbf{\delta x}(\chi\to\chi') +
\mathbf{y}_{\chi'} - \mathbf{y}_{\chi}$, and the total diffusivity is a sum over contributions from every state $\chi$ with probability $P(\chi)$ to any other state $\chi'$,
\begin{equation}
  \kappa(\chi,\chi') := \frac{1}{2d}W(\chi\to\chi')\widetilde{\delta x}^2(\chi\to\chi');
  \label{eqn:kinosons}
\end{equation}
these contributions we call ``kinosons'' (for ``little moves'').%
\footnote{Note also that $dD/d\ln W(\chi\to\chi') = P(\chi)\kappa(\chi,\chi')$, and $P(\chi)\kappa(\chi,\chi')=P(\chi')\kappa(\chi',\chi)$.}
With the optimal $\mathbf{y}_\chi$, the density of kinosons
\begin{equation}
  g(\kappa) :=  \Bigg\langle \sum_{\chi'}\int_0^\infty \delta\left(\kappa-\kappa(\chi,\chi')\right) \Bigg\rangle_\chi
  \label{eqn:density}
\end{equation}
defines the diffusivity for the system, and can elucidate the diffusion process.

Having a variational form in Equation~\ref{eqn:variation} permits a variety of methods to find the optimized displacements, including combining methods for new solutions\cite{TrinkleVariational2018}. For example, using kinetic Monte Carlo to generate finite length trajectories, Equation~\ref{eqn:Einstein} is a variational solution, even if trajectories were to only include a single transition\cite{TrinkleVariational2018}. We can apply parameterized solutions that map our states $\chi$ into vectors $\mathbf{y}_\chi$, such as convolutional neural networks \cite{Cohen2016}, cluster expansions\cite{Nastar2005,Schuler2020}, or a scaling of the average single-transition displacement out of a state (called the ``velocity bias,'' $\mathbf{b}_\chi$ that contains the escape rates from a state). The neural networks contain the linear cluster expansion models as a subset, and should outperform them, while the velocity bias contains explicit information about the rates. For all models, we use machine learning training methods with Equation~\ref{eqn:variation} as our objective function; we construct two equal-sized sets of states with Monte Carlo: a ``training set'' and a distinct ``validation set'' of states. In this case, the ML algorithm \textit{does not have access to the true diffusivity}, but instead optimizes to the lowest diffusivity, and our validation set verify that we have not overtrained our models or have insufficient diversity in the training set. With a variational approach, the algorithm with the lowest diffusivity is closest to the true value, and we can combine multiple methods to increase the accuracy. Finally, once we have our optimized $\mathbf{y}_\chi$, we can find the density of kinosons to identify important diffusion processes.

We apply this new approach to diffusion one of the first and most-studied high entropy alloy, the Cantor alloy\cite{Cantor2004,Zhang2022}; details are available in the supplemental material%
\footnote{See Supplemental Material at http://link.aps.org/supplemental/ 10.1103/PhysRevLett.XXX.YYYYYY for detailed description of parametrics models and objective function, algorithms and dataset construction, and results for lattice gas and high entropy alloys}.
In this equiatomic Mn-Fe-Cr-Ni-Co alloy, diffusion occurs by the movement of a single vacant site (a vacancy) on a face-centered cubic lattice with a random arrangement of five different chemical species. Due to the range of different exchange rates for each species and the dependency on the local environment, complex behavior emerges: fast exchanging species trap a vacancy when the vacancy cannot find a different atom to exchange with to escape. For a face-centered cubic lattice with twelve neighbors, this percolation-like behavior is expected when the concentration of fastest species drops below 20\%\cite{Gaunt1983,Ouyang1989}. Experimental and theoretical investigations of high entropy alloys have identified improved mechanical properties\cite{Gali2013,Wu2014,He2016,Ye2016,Wei2018,Li2018,Fan2022} and high temperature stability which indicates slow kinetics for phase separation\cite{Tsai2013,Choi2018}. The slow kinetics is often surprising, due to the inclusion of ``fast'' exchanging species in the alloys, and presents an intriguing test for the development of accurate theories of diffusion\cite{Thomas2020}.

\begin{figure}
  \includegraphics[width=\figwidth]{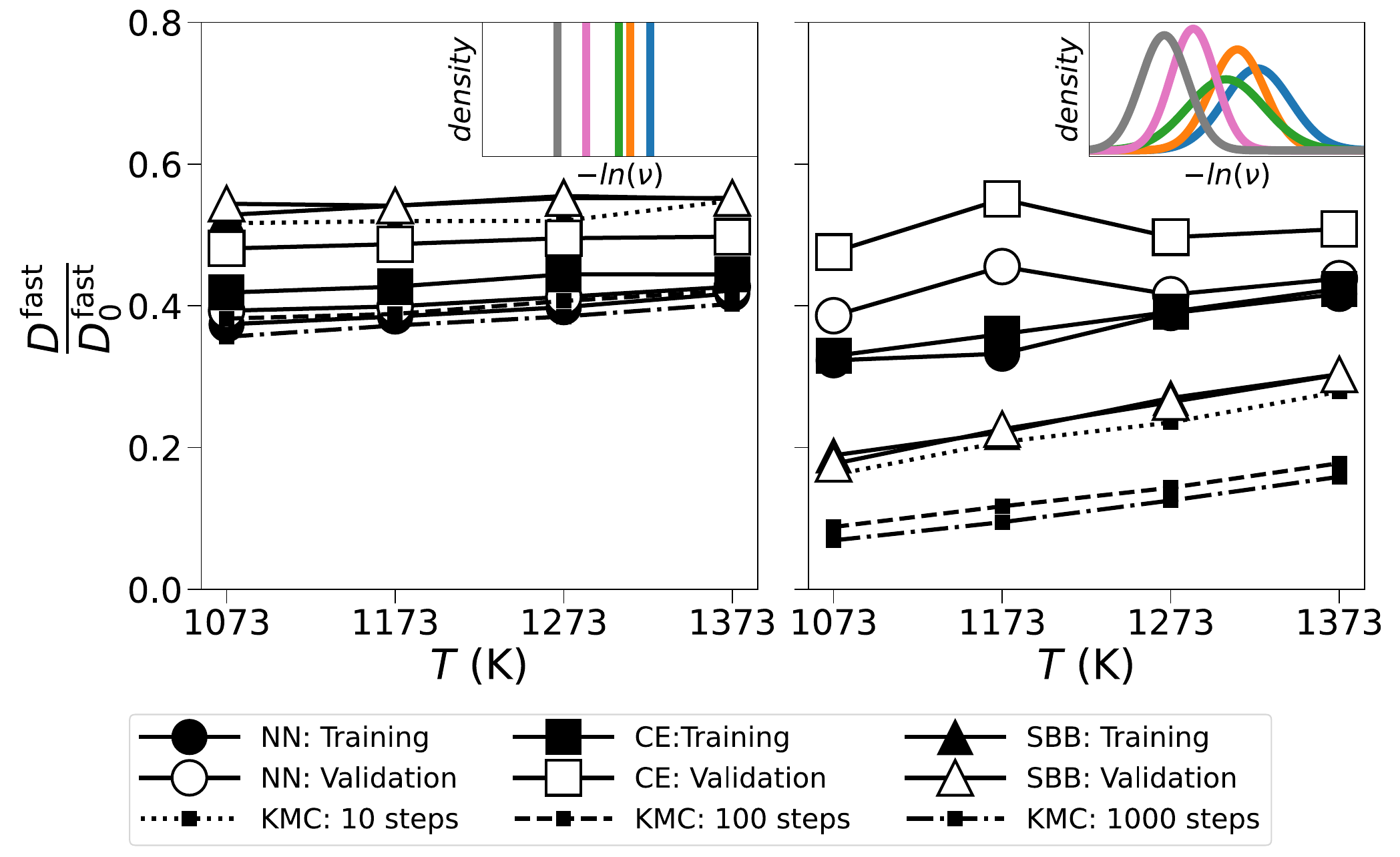}
  \caption{Predicting diffusivity in an ``ordered'' (left) vs. a ``random'' (right) five component model system. Diffusivity is shown as a fraction of the one-step diffusivity prediction; with smaller values indicating increased trapping. The five atomic components have equal concentration, and move via exchange with a single vacancy, either with a fixed rate (ordered) or a distributed rate (random) that depends on the chemistry. The fastest species dominates the exchanges with the vacancy, but has difficulty moving over longer distances as it needs to connect with other fast species to escape trapping. The kinetic Monte Carlo results converge to the true diffusivity as the number of steps increases (lower diffusivity is more accurate). In the ordered case, a neural network (NN) is most accurate, with only a single step; in the random case, the rate-informed scaled-bias basis (SBB) method is the most accurate. A complex system lies in between ordered and random, and we expect needing a combination of methods to predict diffusivity.
  }
  \label{fig:complexity}
\end{figure}

A complex material system lies along a spectrum from ordered to fully random; the two ends of the spectrum provide insight into the performance of different computational methods. We consider two model systems with five components to understand how these approaches behave on the ordered to random spectrum: one where the vacancy exchange rate is fixed for each chemical species, and another where the rate is sampled from log normal distributions convolved with the local environment out to the third nearest neighbor distance of the vacancy.
We select the mean rate (ordered) and mean and variance (random) for the chemical species to match the distributions of our complex high entropy alloy. Figure~\ref{fig:complexity} shows that the ordered system diffusivity can be accurately predicted a nonlinear model (neural network, NN), which outperforms the linear model (cluster expansion, CE), and does better than a simple model relying only on the rates (scaled bias basis, SBB); the configuration around the vacancy provides sufficient information to build an optimized estimate of the mean displacement from a state. However, for a truly random system, the rates are much more diverse, and flummox nonlinear and linear models based on the configuration of species around the vacancy; direct information about the rate is needed to approach what comes from 10 steps in a trajectory. This type of random model has been previously used to estimate diffusivity in high entropy alloys\cite{Thomas2020, Kottke2020, Xu2022_1}, but we note that a random model behaves differently than a \textit{complex} system.

\begin{figure}
  \includegraphics[width=\figwidth]{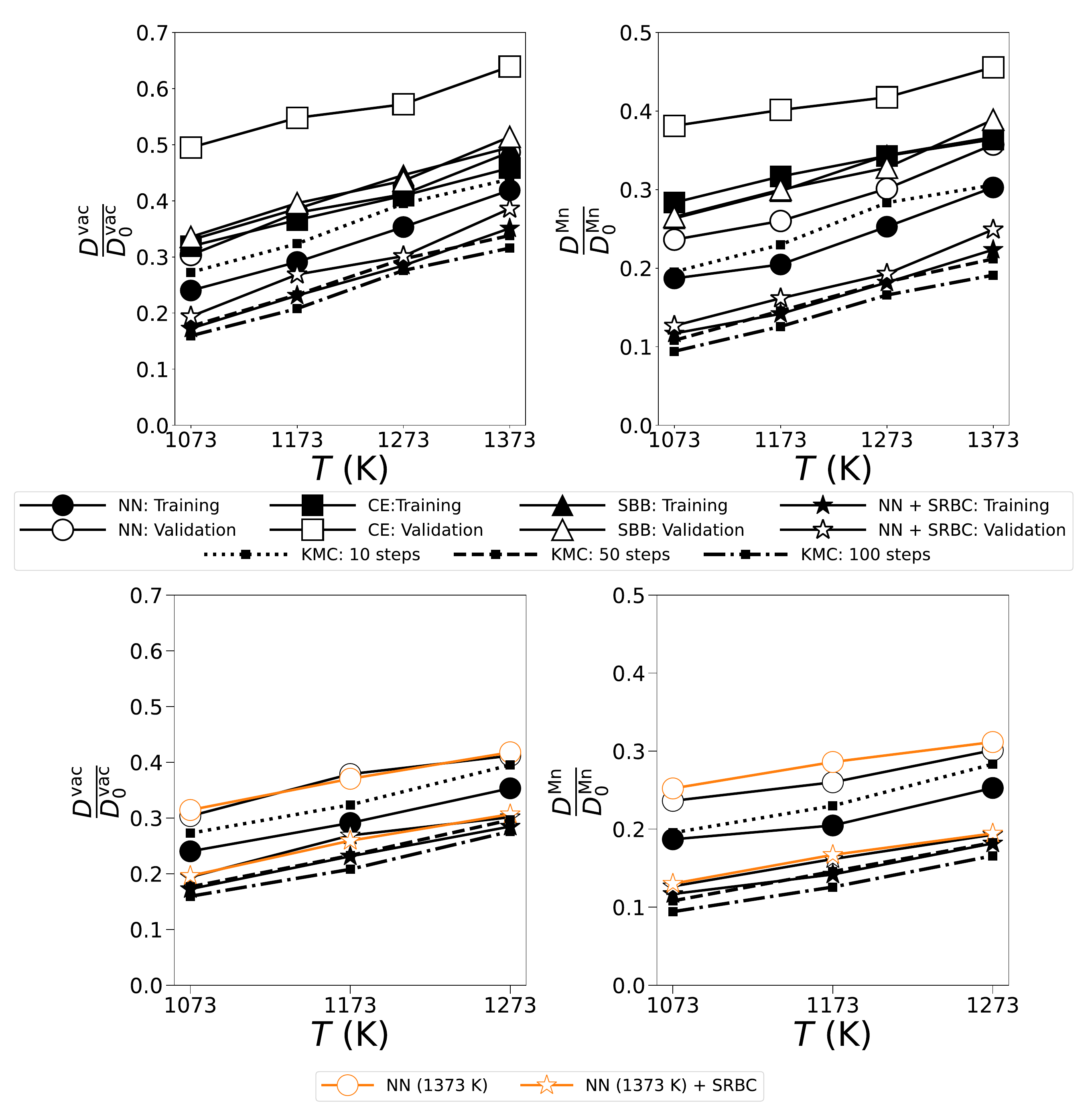}
  \caption{Predicting vacancy (left) and Mn (right) diffusivity in a five component high entropy alloy. Diffusivity is shown as a fraction of the one-step diffusivity prediction; with smaller values indicating increased trapping. The real alloy has atomic exchanges with a vacancy that depend on the local atomic environment, behaving as a combination of an ordered and random system. The kinetic Monte Carlo results converge to the true diffusivity as the number of steps increases (lower diffusivity is more accurate). By combining a neural network with a rate-informed scaled residual bias correction (NN+SRBC, stars), it is possible to predict diffusivity very close to the true value, while only taking a single transition from one state to another. The neural network trained on higher temperature states can work at lower temperatures in the same phase, with similar accuracy (orange stars). This accuracy allows the computation of the density of kinosons for this complex alloy, and an understanding of the diffusion process.
  }
  \label{fig:HEA}
\end{figure}

A complex high entropy alloy combines aspects of both ordered and fully random systems, and so we combine a convolutional neural network with a scaled residual bias correction in Figure~\ref{fig:HEA}. The diffusivity is converged after 100 transitions in a trajectory using Equation~\ref{eqn:Einstein}, or we can instead use a ML approach to get similar results with \textit{single} transitions and Equation~\ref{eqn:variation}. The ML approach starts with a neural network to transform local environments around the vacancy into an estimate for $\mathbf{y}_\chi$ by minimizing the diffusivity in Equation~\ref{eqn:variation}; we can further correct that estimate with the residual velocity bias by directly incorporating information about the escape rates. We note that the neural network can be optimized using high temperature kinetics, and applied out-of-domain at lower temperatures. This combined method outperforms other computational approaches to the diffusivity, while also requiring orders of magnitude less computational effort than computing long-time trajectories. As we have a good approximation for our optimized displacements, we also evaluate the kinosons for this material system.

\begin{figure}
  \includegraphics[width=\figwidth]{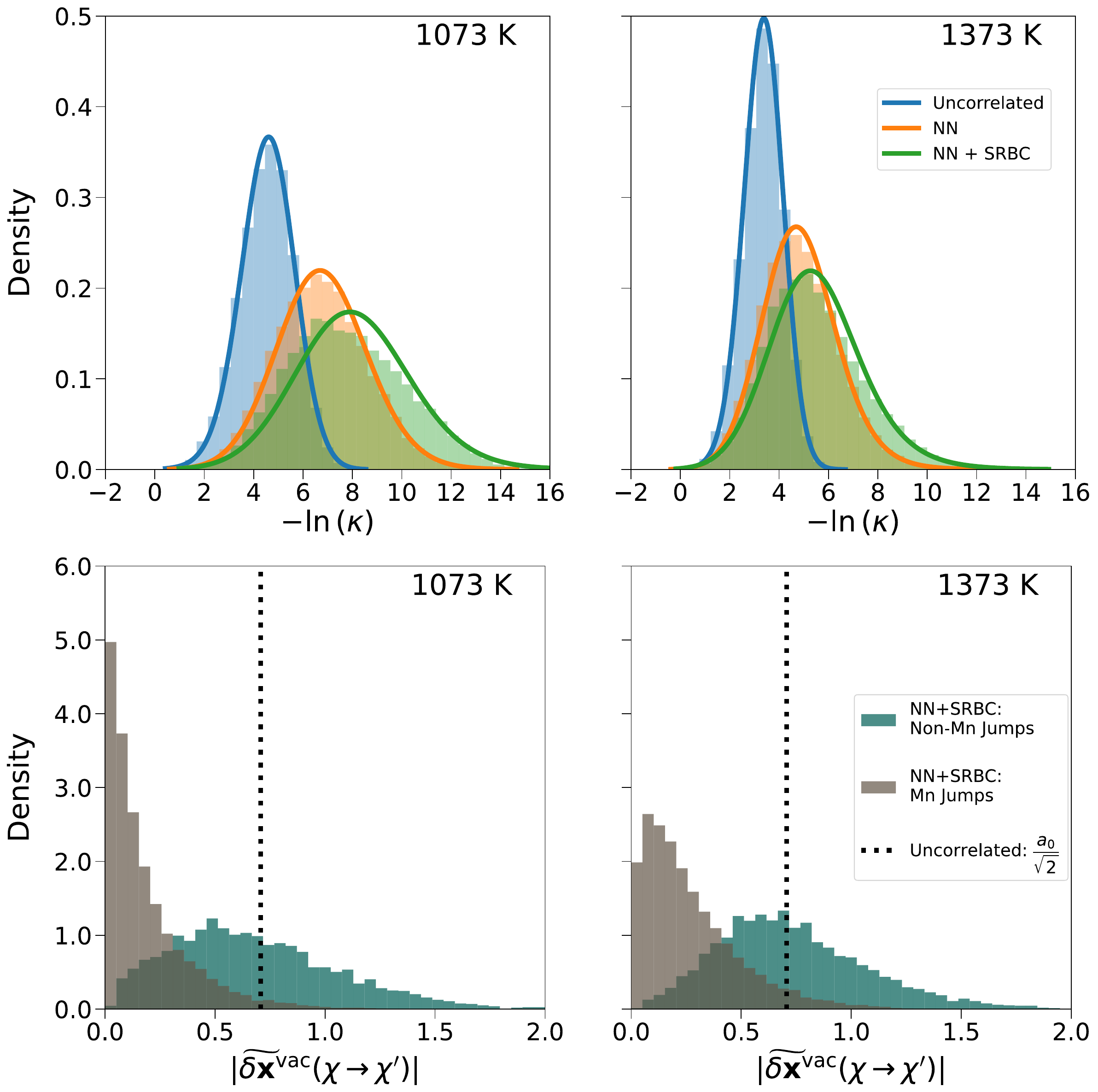}
  \caption{Density of kinosons $\kappa$ (top) and optimized displacements $\widetilde{\mathbf{\delta x}}$ (bottom) for the vacancy in a high entropy alloy. Without the optimized displacements, the distribution of exchange rates provides an ``uncorrelated'' estimate of the density of kinosons that contribute to diffusivity; for this alloy, this is a nearly log-normal distribution. Once the displacements are known, the kinoson distribution skews to lower values, and follows a log exponentially-modified Gaussian. The optimized displacements show a distinct distribution depending on whether they are with a fast Mn or one of the slower elements; the Mn displacements collapse to a peak at zero, while the other elements are distributed around the jump distance in the face-centered cubic lattice. This change is a signature of percolation-like behavior for Mn.
  }
  \label{fig:kinosons}
\end{figure}

Figure~\ref{fig:kinosons} shows how the density of kinosons differs from the distribution of rates in the problem, with a fingerprint of percolation of the fast species in the optimized displacements. The complex high entropy alloy has five different normal distributions for the energy barriers to exchange with a vacancy, depending on the chemical species; the distribution of rates is then approximately log-normal and dominated by the fastest species (Mn). However, the density of kinosons follows a different distribution: a log-skewed normal, which we identify with an exponentially modified Gaussian form\cite{Grushka1972, Olivier2010, Ali2022}. The mean of $\ln \kappa$ moves to lower values, while the skewness also favors smaller values. The change in form is due to the relaxed displacements $\widetilde{\delta x}$, which also show their own interesting behavior. The vacancy exchanges with all five species in the alloy, albeit with highest probability for Mn. However, the distribution of $\widetilde{\delta x}$ looks remarkably different for Mn---where it collapses to highest probability at \textit{zero displacement}---than the other four species. While Mn is the fastest exchanger, at 20\%\ concentration most exchanges are later undone, and hence the maximum probability for zero displacement. The collapse of the distribution indicates the percolation limit, without connected paths for Mn to transit.

\begin{figure}
  \includegraphics[width=\figwidth]{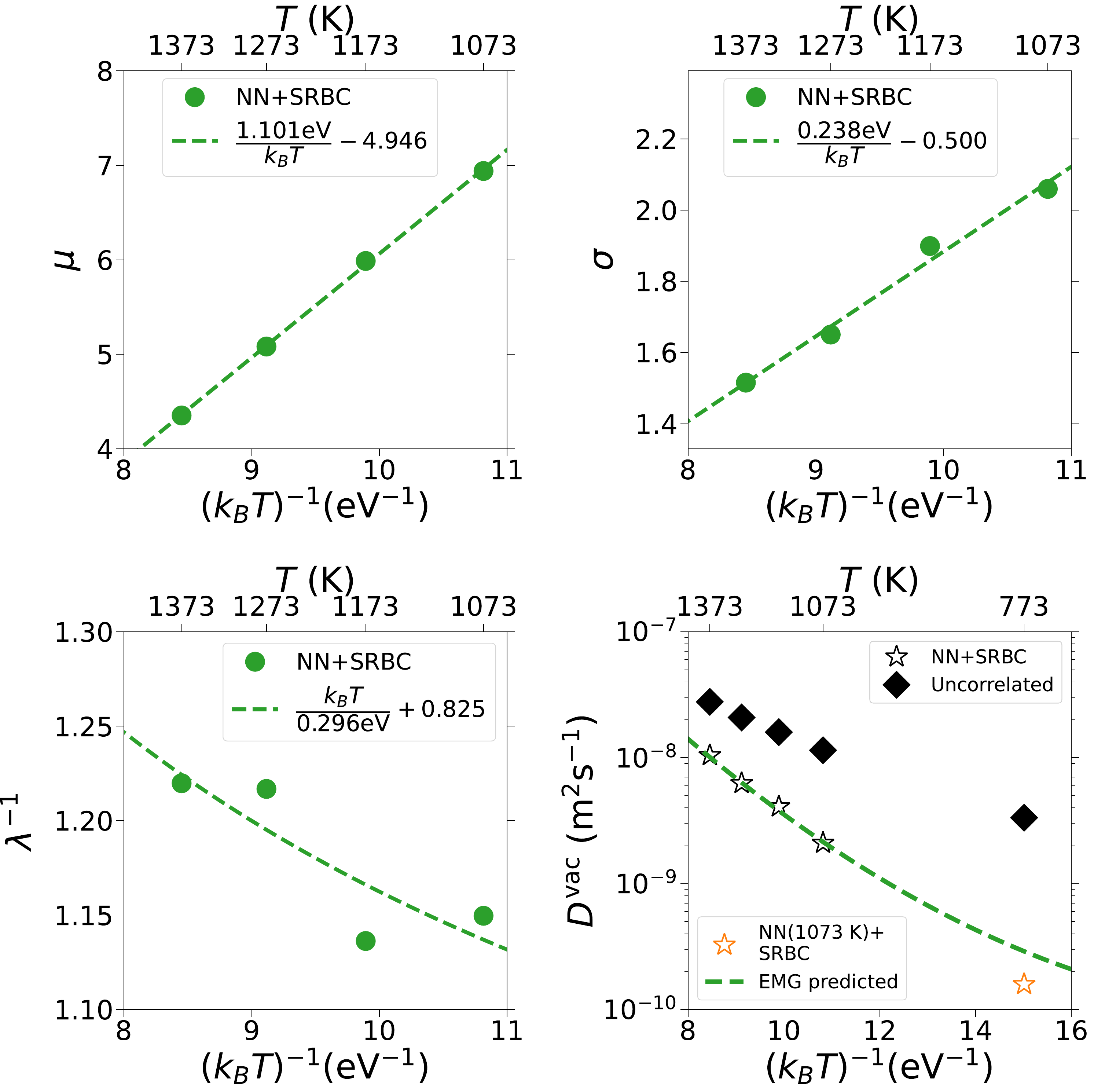}
  \caption{The density of kinosons's mean, standard deviation, and decay parameters with temperature for the high entropy alloy, and extrapolation to lower temperatures. The log-exponentially modified Gaussian form in Figure~\ref{fig:kinosons} is quantified with three parameters, and their temperature dependence empirically fit. This reveals a new non-Arrhenius analytic form for the vacancy diffusivity (Equation~\ref{eqn:EMG}), that can be extrapolated accurately to lower temperatures.
  }
  \label{fig:EMG_params}
\end{figure}

Finally, we can examine the behavior of the density of kinosons with temperature, and derive a new functional form for the diffusivity of high entropy alloys in Figure~\ref{fig:EMG_params}. An exponentially modified Gaussian distribution has three parameters: mean $\mu$, variance $\sigma^2$, and decay parameter $\lambda^{-1}$ that controls the skewness. The parameters follow simple temperature behavior that allows us to easily fit the macroscopic diffusion behavior, and extrapolate to lower temperatures. 
The change in the density of kinosons with temperature can be modeled with a few parameters, producing a new functional form for this diffusivity in this class of alloys. In terms of our parameters, the diffusivity is
\begin{equation}
    D^\mathrm{(EMG)} = \frac{\lambda}{\lambda + 1}\exp\left(-\mu + \frac{1}{2}\sigma^2\right)
    \label{eqn:EMG}
\end{equation}
If $\sigma$ and $\lambda^{-1}$ were both zero, the diffusivity would follow an Arrhenius form as $\mu$ is linear in inverse temperature; however, $\sigma$ is also linear in inverse temperature and $\lambda^{-1}$ is finite with a weak temperature dependence, producing \textit{non-Arrhenius behavior} in this complex high entropy alloy. We note that the vacancy and Mn have non-zero $\lambda^{-1}$ while Fe and Cr are better represented by log-normal distributions ($\lambda^{-1}=0$). The result of this high temperature fit also extrapolates well to even lower temperatures. The functional form of Equation~\ref{eqn:EMG} comes from the density of kinosons, revealed by our machine-learning analysis, and we expect it to be applicable to other high entropy alloys.

\begin{figure}
  \includegraphics[width=0.75\figwidth]{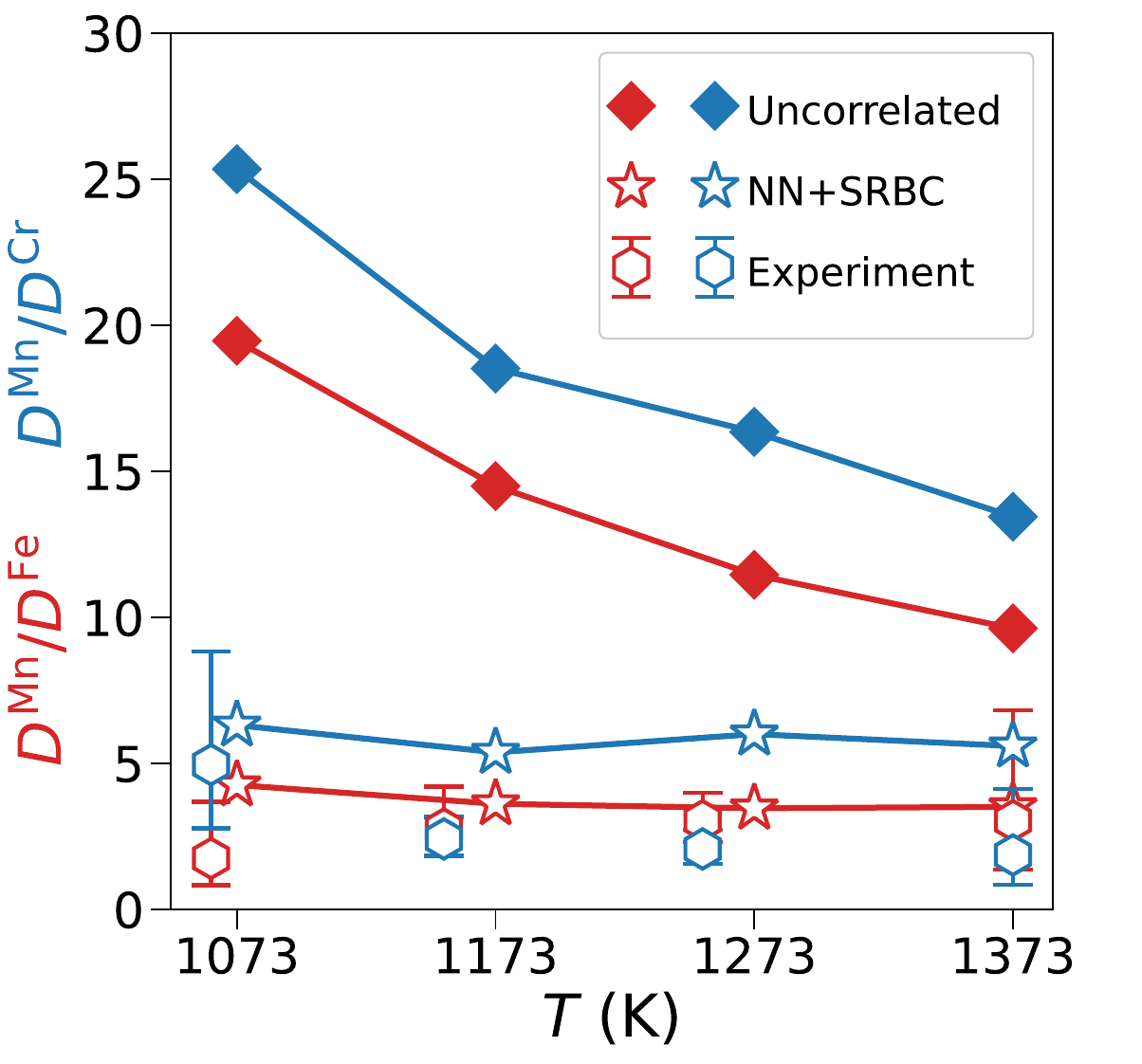}
  \caption{Ratio of diffusivity of Mn to Fe and Cr. Using the average rates (``uncorrelated'') disagrees with experimental tracer diffusion measurements\cite{Gaertner2020}, while the neural network calculations of the optimized kinosons are within experimental uncertainty.}
  \label{fig:Experiment_compare}
\end{figure}

The new prediction of diffusivity agrees much better with experimental measurements\cite{Vaidya2018} in Figure~\ref{fig:Experiment_compare}. The ratio of diffusivity for the second and third fastest species (Fe and Cr) to the fastest (Mn) removes the unknown vacancy concentration, tracer correlation factor, and thermodynamic factors in our random alloy---assumed to be equal for our species---to robustly compare with the measured tracer diffusivity from experiments. The unrelaxed kinoson prediction of ``uncorrelated'' diffusivity---what you would expect using the distribution of transition rates---shows a significant deviation from the experimental values. When the optimized kinosons are used instead, the agreement is within the experimental error estimates.

Working with a variational approach for diffusivity, a machine-learning approach for optimization correctly predicted diffusion in a complex high entropy alloy with a fraction of the effort needed for trajectories. We take advantage of a physics-informed neural network, leveraging underlying crystal symmetries including translation and the locality provided by a convolutional neural network to gain insight into the underlying physical processes. With the optimized displacements, we express diffusivity as a sum of discrete jumps, called kinosons, which follow a different statistical distribution than the transition rates in the material. The distribution of optimized displacements indicates the approach of a percolation transition for the fastest species in alloy, helping to illustrate the complexity of this material that lies between ordered and random. The density of kinosons provides a new analytic form for diffusivity of high entropy alloys, and the parameters can be easily fit from the density of kinosons. This new analysis technique highlights the power of physics-based machine-learning to model and even \textit{understand} diffusion processes in complex materials, transforming the high-dimensional kinetic problem of diffusion into a one-dimensional density of kinosons. This should prove a powerful tool for understanding diffusion in other solid materials, including glasses, and analysis of long-time kinetic processes may be applied to other nonequilibrium problems in materials or chemistry.

\nocite{Trinkle2018,Trinkle2017,glorot,Torch2019,Adam,Ebihara2017,Allnatt2016,Manning1971,Moleko1989,frenkel2023,lammps,Nounou2000,Henkelman2000,E2002,Maras2016,Restrepo2023,Gunol2020,Garnier2013,Porter2021_chp_1,Sugita_2020}

\begin{acknowledgments}
The authors thank Dr.\ Danny Perez and Prof.\ Lee DeVille for helpful conversations, and Prof.\ Sergiy Divinski for helpful conversations and the experimental data in Fig.~\ref{fig:Experiment_compare}.
This work is sponsored by the NSF under program MPS-1940303. This work made use of the Illinois Campus Cluster, a computing resource that is operated by the Illinois Campus Cluster Program (ICCP) in conjunction with the National Center for Supercomputing Applications (NCSA) and which is supported by funds from the University of Illinois at Urbana-Champaign.
The code is available at https://github.com/TrinkleGroup/VKMC, and the data is available at
https://zenodo.org/doi/10.5281/zenodo.10214333.
\end{acknowledgments}


%
\newpage
\includepdf[pages={{},-}]{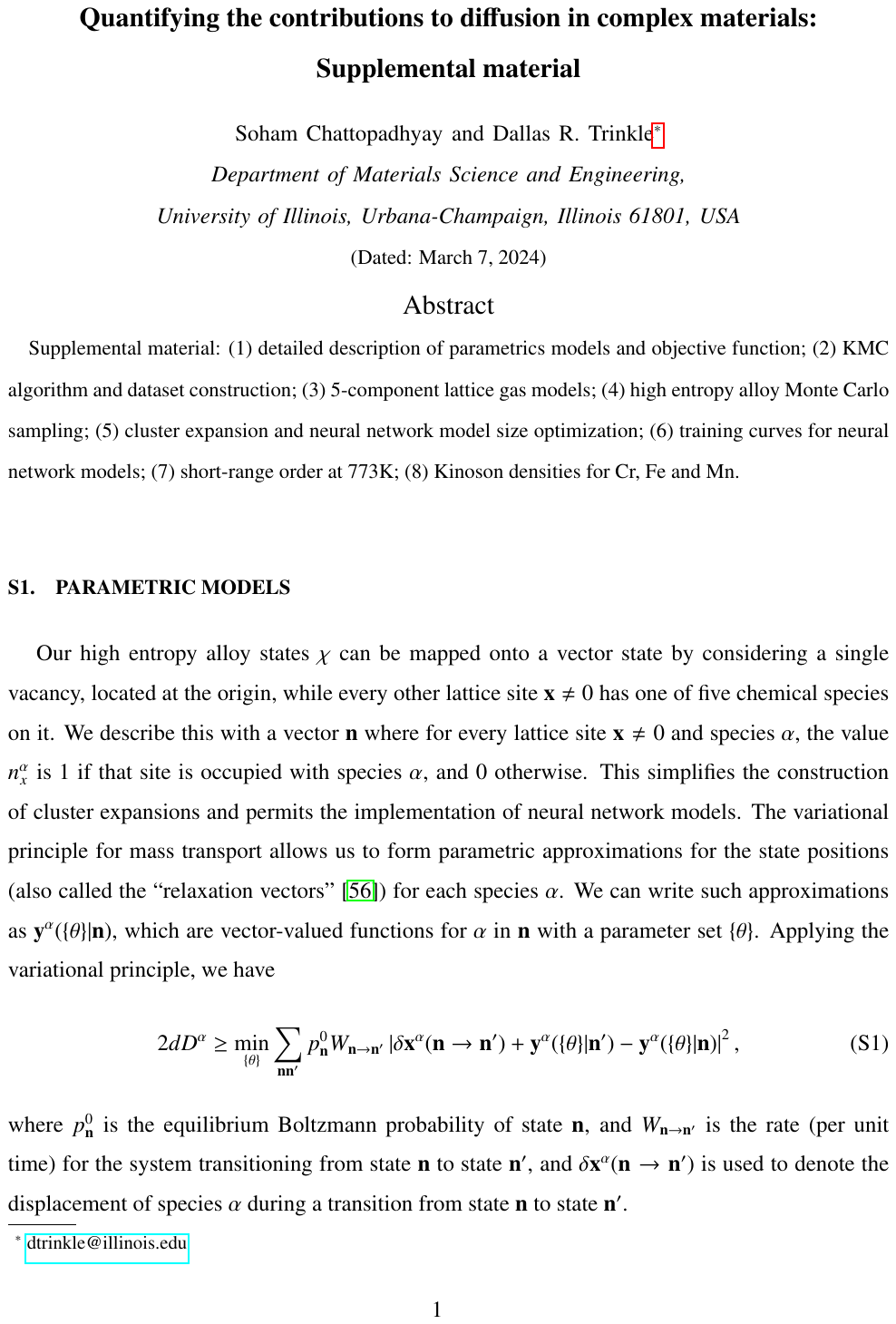}
\end{document}